\documentclass[10pt,conference]{IEEEtran}

\usepackage{booktabs}
\usepackage{tabularx}
\usepackage{amsmath}
\usepackage{cite}
\usepackage{balance}
\usepackage{graphicx}
\usepackage{array}
\usepackage{xcolor}
\usepackage{xurl}  
\usepackage[hidelinks]{hyperref}

\makeatletter
\newcommand{\linebreakand}{%
\end{@IEEEauthorhalign}
\hfill\mbox{}\par
\mbox{}\hfill
\begin{@IEEEauthorhalign}
}
\g@addto@macro\normalsize{%
  \setlength\abovedisplayskip{3pt plus 1pt}%
  \setlength\belowdisplayskip{3pt plus 1pt}%
  \setlength\abovedisplayshortskip{1pt plus 1pt}%
\setlength\belowdisplayshortskip{2pt plus 1pt}}
\makeatother

\newcolumntype{L}{>{\raggedright\arraybackslash}X}
\newcolumntype{R}{>{\raggedleft\arraybackslash}X}

\usepackage[most]{tcolorbox}
\definecolor{answerbg}{HTML}{EEF4FB}
\definecolor{answerrule}{HTML}{2A78D6}
\newtcolorbox{rqanswer}[1]{enhanced, colback=answerbg,
  colframe=answerrule, boxrule=0pt, leftrule=2.5pt, arc=0pt, outer arc=0pt,
  left=5pt, right=5pt, top=4pt, bottom=4pt, before skip=8pt, after skip=8pt,
  fontupper=\small, title={Answer to #1}, coltitle=answerrule,
fonttitle=\bfseries\small, attach title to upper={\ }}

\begin{document}

\title{The AI Era's Lines of Code? A Case Study on What an Engineer's Token
Spend Measures}



\author{
  \IEEEauthorblockN{Shyam Agarwal}
  \IEEEauthorblockA{Carnegie Mellon University\\
  shyamaga@andrew.cmu.edu}
  \and
  \IEEEauthorblockN{Olga Lvova}
  \IEEEauthorblockA{JetBrains\\
  olga.lvova@jetbrains.com}
  \and
  \IEEEauthorblockN{Agnia Sergeyuk}
  \IEEEauthorblockA{JetBrains Research\\
  agnia.sergeyuk@jetbrains.com}
  \and
  \IEEEauthorblockN{Anastasiia Serova}
  \IEEEauthorblockA{JetBrains\\
  anastasiia.serova@jetbrains.com}
  \linebreakand
  \IEEEauthorblockN{Timofey Bryksin}
  \IEEEauthorblockA{JetBrains Research\\
  timofey.bryksin@jetbrains.com}
  \and
  \IEEEauthorblockN{Thomas Zimmermann}
  \IEEEauthorblockA{University of California, Irvine\\
  tzimmer@uci.edu}
  \and
  \IEEEauthorblockN{Bogdan Vasilescu}
  \IEEEauthorblockA{Carnegie Mellon University\\
  vasilescu@cmu.edu}
}

\maketitle

\begin{abstract}
  Organizations use engineers' token spend to monitor AI adoption, set budgets,
  and measure productivity, but the metric's relationship with engineering output is
  unclear. We link five weeks of token usage in coding agents to the commits of
  404 engineers at JetBrains, and analyze
  a company survey of 412 employees. We find that token spend varied mostly between
  engineers, and which engineers spent the most stayed largely the same over
  the five weeks. Among weeks with similar commit counts, the 90th percentile
  of spend was 39 to 240 times the 10th. Within an
  engineer, doubling spend was associated with about 10\% more commits.
  Available work metrics explained 38\% of the variation in spend between
  engineers and 13\% of the week-to-week variation within an engineer. In the
  survey, respondents
  in technical roles reported in nearly nine of ten agent evaluations using
  agents for tasks beyond
  coding, such as analysis, research, and planning. Token spend
  therefore measures how much engineers use agents across many kinds of
  work and not the outcome and value of that work.
  We recommend using token spend at most for resource
  accounting, and evaluating agent-assisted work with evidence about tasks
  and outcomes.

\end{abstract}


\pagestyle{plain}

\section{Introduction}
\label{sec:intro}
Many engineers now write software by instructing a coding agent, reviewing
what it did, and
asking for changes or further
work~\cite{liu2026copilot,he2026,murphyhill2026}. Whatever code these interactions
produce is still recorded, as before, in commits and pull requests.
Every exchange with the agent is also recorded in the agent's usage log,
or \emph{agent telemetry}. For each exchange, the model provider
logs and bills the tokens sent and received, along with the tool and model
used. Summed per engineer and per week, these counts are the engineer's
\emph{token spend}. Spend counts every exchange, including those that never
lead to a commit, but it does not record what an exchange was about.
Providers do not share session content by default, and few companies record
it themselves, so the organization that pays usually sees only the token count, as
in our data (Section~\ref{sec:data}).

Some organizations already use token spend to evaluate engineers. At Meta, an employee
built a dashboard on which colleagues competed to be the company's top token
user; it was taken down after the press reported on
it~\cite{fortune2026meta,bilicarar2026}. Executives at Nvidia and Meta have
said that a good engineer should use tokens worth a large share of their
salary~\cite{nieva2026forbes}, and Salesforce has set minimum
monthly spending targets~\cite{orosz2026tokenmaxxing}.
Analytics vendors divide token cost
by merged pull requests and report the ratio as return on
investment~\cite{jellyfish2026tokenmaxxing,weave2026index}. Each of these
uses assumes that an engineer who spends more produces more. Practitioners
have begun to doubt that assumption. Uber spent its 2026
budget for AI coding tools in four
months; its COO said that the link between token spend and shipped
features ``is not there yet''~\cite{bi2026uber}. One critic called token
spend ``the lines-of-code metric of the AI era: easy to measure, easy to
game, and disconnected from actual productivity''~\cite{bilicarar2026}.

\begin{figure}[t]
  \centering
  \includegraphics[width=0.85\columnwidth]{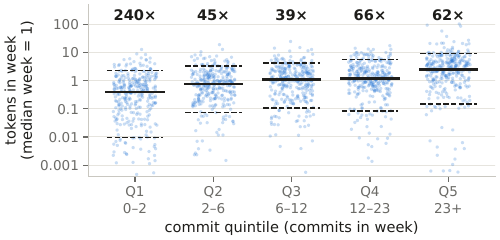}
  \caption{Within each weekly-commit quintile, the 90th percentile of token
    spend is 39 to 240 times the 10th (numbers at the top). Lines mark the 10th, 50th,
    and 90th percentiles; spend is relative to the median engineer-week.
  Ties are split at quintile edges, so adjacent ranges share an endpoint.}
  \label{fig:spread}
\end{figure}

\looseness=-1
Prior studies of whether more tokens mean more output give mixed answers. On
benchmarks, agents that run longer do not reliably
solve more tasks~\cite{bai2026spend,fan2025sweeffi,wu2026tokenmaxxing}.
In deployed use, output rises far less than spend: a press summary of vendor
data reports that developers who spend ten times more merge about twice as
many pull requests~\cite{fernholz2026tokenmaxxing}, and one
enterprise doubled its throughput but raised its spend at least
twenty-fold~\cite{he2026}. Outside benchmarks, these comparisons are
between engineers, or between organizations. Engineers differ in
role, team, and work practices, so a difference in spend between two of them may
reflect those differences as much as any difference in what they produced.
No study has followed both spend and output for the same engineers over
time, so whether an engineer's spend varies with their output from
week to week is not known. Telemetry from one agent shows how widely usage
varies across users but does not have output data to compare it
with~\cite{liu2026copilot}, and recent panel studies measure agent use by
days~\cite{murphyhill2026}, interactions~\cite{heilman2026}, or lines the
agent wrote~\cite{he2026}, and report spend only in aggregate, if at all.

In this paper, we ask what an engineer's token spend measures and what it
can be used for. We study 404 engineers at a software development company
whose internal gateway meters the tokens of every coding-agent call. From
the gateway counts and the company's repositories, we follow each
engineer's token spend and commits week by week for five weeks, and we
analyze an anonymous survey of the company's
employees about their AI use.


We ask two questions. \textbf{RQ1:} \emph{How closely is an engineer's
token spend related to their output?} We compare an engineer's weeks with
one another, which holds the engineer fixed, and we compare engineers with
each other. \textbf{RQ2:}
\emph{Where does the token spend that is not tied to repository output go?}
That spend may be work that never becomes a commit, which studies of
repository output would miss, or overhead. The
telemetry shows when each call happened and how many tokens it
used, but not what it was for, so it cannot distinguish the two. We use the telemetry to
estimate the share of spend not tied to output, and the survey to learn
what engineers use agents for and how many hours they report saving.

We find that spend depends more on the engineer than on the week's output (RQ1).
A week with twice an engineer's usual spend has only about 10\% more commits,
weeks with similar commit counts differ in spend by a factor of 39 to 240
(Figure~\ref{fig:spread}), and the engineers who spend the most are
largely the same over the five weeks. Up to about two fifths of spend does
not vary with the week's
output, and in 89\% of technical respondents' evaluations of a coding agent,
they had used it that week for something other than writing code (RQ2).
Engineers choose how much to use agents, so these are associations.

Our contribution is an empirical study that links engineers' metered token
spend to their repository output and to their own accounts of agent use. It
shows that token spend measures how much an engineer works through agents,
not what they produce, so it should not be used as a measure of productivity.

\section{Related work}
\label{sec:related}

Organizations started adopting token spend as a management metric before
peer-reviewed research could examine it, so we follow multivocal review
guidance~\cite{garousi2019multivocal} and include grey literature (vendor
reports, practitioner accounts, and technical press) alongside academic
work.

\subsection{Measuring developer productivity}
\label{sec:rw-metrics}

Lines of code became a productivity measure because they were cheap to
count, and they have long been criticized for penalizing expressive
languages and inviting padding~\cite{albrecht1979,jones1986,fenton1997,kaner2004}.
Still, they stayed in use because program size is associated with effort closely
enough to support cost models~\cite{fenton1997}. That history suggests that
an indicator that is convenient to collect is not automatically a valid proxy
for the attribute of interest~\cite{kaner2004}, and that a measure that
becomes a target ceases to be a
good measure~\cite{strathern1997improving}. SPACE~\cite{forsgren2021} and
DORA~\cite{forsgren2018accelerate}, which replace single numbers with
several, measure outputs, outcomes, and
processes, but not the resources used, a reasonable omission while the
main input was human time. For DX Core 4, which builds on
both, AI makes the metrics noisier, because ``the same number can now be
produced by very
different combinations of human and AI behavior''~\cite{houck2026}. Empirical
studies of AI-assisted development
mostly observe work through commits, pull requests, and
reviews~\cite{agarwal2026ides,denisovblanch2026markdown,he2026,murphyhill2026}.
Agent telemetry adds a record of the resources used, metered per person,
including the interactions that don't produce any artifact.

\subsection{Task-level evidence}
\label{sec:rw-task}

Most academic work on token consumption studies agent trajectories on
benchmarks such as SWE-bench~\cite{jimenez2024swebench}. Accumulated context and
cache reads, not generated output, account for most tokens and most
cost~\cite{bai2026spend,salim2026tokenomics,weinberger2026token,liu2026copilot}.
Spending more does not reliably increase success: accuracy peaks at
intermediate cost, failed attempts cost more than successful ones, and the
extra tokens go to repeated file views, retry loops, and generation after
a passing solution~\cite{bai2026spend,fan2025sweeffi,wu2026tokenmaxxing,liu2026copilot}.
Much of the variance is unrelated to the task: consumption depends on
model, scaffold, and language, and only weakly on expert-rated
difficulty~\cite{bai2026spend,fan2025sweeffi,wu2026tokenmaxxing}.

Token counts are also an unstable proxy for cost. Since cost became an
evaluation dimension~\cite{kapoor2024agents}, many methods reduce
tokens~\cite{gao2026turn,guo2026eet,xiao2026trajectory,solovyeva2026babbling,sun2025tokensugar,pan2026readability},
but fewer tokens do not always mean lower cost. Removing 38.4\% of
tool-output tokens \emph{increased} billed cost by 6.8\%, because the agent
spent extra turns diagnosing and re-reading the removed
context~\cite{weinberger2026token}. With feature, review gate, and model
held constant, cost per merged feature varies about 2.5-fold across
harnesses~\cite{vanamo2026}. If token counts are an unreliable measure of
cost at the task level, they are unlikely to be a reliable measure of value
at the organizational level.

\subsection{Individual- and organization-level evidence}
\label{sec:rw-org}

We did not find any peer-reviewed study of how token spend relates to engineering
outcomes across developers or organizations. Telemetry studies show that
spend varies 50-fold or more between users but do not report
outcomes~\cite{liu2026copilot,tei2026}, and a single-developer case study
links higher token use to a lower fix-commit ratio but changed model and
harness at the same time~\cite{peng2026inference}. Adoption studies find more commits or
merged pull requests after adoption, with the gains shrinking before
release~\cite{agarwal2026ides,denisovblanch2026markdown,murphyhill2026,he2026cursor,demirer2026},
but they compare use with non-use and do not measure tokens. Two panel studies
closest to ours measure how intensively engineers use AI tools,
though also not in tokens. He et
al.\ count AI-written lines and find that a mandate to double merged pull
requests raised throughput 2.09-fold while spend increased at least
20-fold~\cite{he2026}; Heilman et al.\ count Copilot interactions and find
40.5\% more completed pull requests in an engineer's weeks of heaviest use
than in weeks without use~\cite{heilman2026}.
Self-reported speedups can diverge from measured time: in a randomized trial,
developers took 19\% longer with AI tools but thought they were
faster~\cite{becker2025metr}.

Only analytics vendors have joined token telemetry to delivery data across
organizations, and they report that throughput rises far less than
spend. From the lowest to the highest usage decile, weekly pull requests
nearly tripled while cost per merged pull request rose about
300-fold~\cite{jellyfish2026tokenmaxxing}, and merged output per dollar fell
by about 70\% during 2026~\cite{weave2026index}. Alongside the throughput gains,
other reports find more bugs and churn, and lower acceptance once later
revisions are counted~\cite{faros2026tokenmaxxing,fernholz2026tokenmaxxing}.
None of these reports is peer reviewed, and their methods are only partly
disclosed.

Practitioners also disagree on what token spend is. Some treat it as a measure of
engagement, equating good engineering with heavy use~\cite{nieva2026forbes} or
setting a daily token threshold for adoption~\cite{bi2026yegge,venturebeat2026}.
Others treat it as a cost to cap~\cite{constantin2026tokenminimizing} or to
divide by output, or as a proxy for output itself, a practice called
\emph{tokenmaxxing}~\cite{bilicarar2026}; one practitioner observes that lines
of code now correlate highly with token spend~\cite{arcolano2026}. Reports of
employees inflating their
usage~\cite{fortune2026meta,bilicarar2026,cooter2026meta,palmer2026amazon}
fit Goodhart-style
gaming~\cite{strathern1997improving} but remain anecdotal.

\subsection{Synthesis and open questions}
\label{sec:rw-synthesis}

Prior work has not answered two questions about the engineers behind the
telemetry. The first is what an individual's spend reflects. Studies that join spend to
output compare deciles of users or follow
organizations~\cite{jellyfish2026tokenmaxxing, weave2026index, he2026}, so
they cannot separate how much an engineer habitually works through agents
from how their use changes with the work at hand. The second is how much
agent-assisted work ends up in the repository. Half of production agent turns
are exploration or reasoning, not editing~\cite{liu2026copilot}, yet
adoption studies measure outcomes through repository
artifacts~\cite{agarwal2026ides, he2026} and look for lost gains only after
the first commit~\cite{demirer2026}; no study has estimated how much consumption
never leads to a commit. Token counts depend on model, harness, and caching
regime~\cite{bai2026spend, weinberger2026token, peng2026inference}, so our
person-week panel addresses both questions within a single tooling
environment, where the counts are comparable.

\section{Methods}
\label{sec:methods}

We combine gateway telemetry of agent use at JetBrains with repository records
of the same engineers' output. We also draw on an anonymous internal survey on
AI tools that the company ran during the study period; survey responses
cannot be linked to individual engineers' telemetry or repository data.
Because engineers choose
how much to use the agents, all our estimates are associations, not causal effects.

\subsection{Setting and data}
\label{sec:data}
\label{sec:missing}

JetBrains builds software development tools and had about
2,600 employees at the time of the study.\footnote{About
JetBrains~\url{https://www.jetbrains.com/company/}} We study coding-agent use
as recorded by the company's internal gateway,
which logs every agent call in a consistent format.

We extracted pseudonymized data from internal systems on the 584
engineers who committed to one of the company's main IDE products in our
study window, 2026-05-10 to 2026-08-08. The commit data spans the
window, but our access to the gateway's
token data spans only its last weeks; we analyze the five complete weeks
in that period, 2026-06-29 to 2026-08-02 (the \emph{analysis weeks}). We
exclude the final six days (2026-08-03 to
2026-08-08), because
commits during that period may have been merged after the extract date.
The seven weeks before the analysis weeks (the
\emph{pre-period}) have commit data but no token data. Engineers had access to AI
tools in the pre-period too, so it is not
a period without AI. We use the pre-period data to describe each engineer's
prior work mix
(Section~\ref{sec:panel}).

\begin{figure}[t]
  \centering
  \includegraphics[width=0.85\columnwidth]{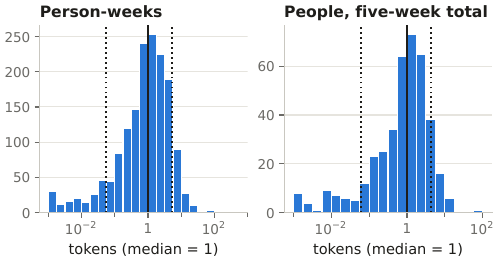}
  \caption{Token spend relative to the median, per metered person-week
    (left, $n = 1{,}599$) and per engineer (right, $n = 397$), over the five
    analysis weeks. Solid lines: medians. Dotted lines: 10th and 90th
    percentiles, at 0.05 and 5.3 times the median for person-weeks and 0.06
    and 4.3 for engineers. Values under a thousandth of the median are pooled
  in the first bin.}
  \label{fig:spend-dist}
\end{figure}

From the company directory we extract each engineer's team, job title, and manager
flag. We assign each engineer a role (developer, QA engineer, team lead or
manager, intern, other) from keywords in the job title. QA and developer keywords take
precedence over lead and manager, so a lead QA engineer is a QA engineer.
The agent telemetry
contains 3.1 million agent calls by 406 of the 584 engineers from
2026-06-29 to 2026-08-08. Each call records a timestamp, the agent (two
  agents account for nearly all tokens: Agent A for 69\% and
Agent B for 31\%), the
model, the number of tokens, and the credits charged.
Most tokens are context re-sent on every turn: the median call sends about
400 times as many tokens as it receives. We measure spend in tokens; credits, the
company's cost unit, rank engineers almost identically (Spearman $\rho =
0.97$) and serve as a sensitivity check. Tokens measure only the inference
part of the input; the engineers' own time, the larger cost, is not in our
data. Spend varies widely between
engineers (Figure~\ref{fig:spend-dist}): over the five analysis weeks, the
engineer at the 90th
percentile used 71 times as many tokens as the engineer at the 10th.

For output we measure the number of commits. We date each commit by its author date,
when the change was written and the agent was used, not by the commit date,
which rebases and squash merges change to be the merge time. For each commit we
also know its size (lines added and deleted and files changed, recorded in
bins; we use bin midpoints), whether it changes the shared IDE platform or a
surrounding project, and the issues referenced in its message, with the
issue's type (bug, task, feature) from the issue tracker. In the analysis
weeks, 71\% of commits change the platform and 35\% reference an issue.
Commit counts miss code quality and
work that does not change the code, such as review and design. We also
considered pull requests, as some prior work
did~\cite{he2026,jellyfish2026tokenmaxxing}, but lacked the data.

\subsection{The telemetry panel}
\label{sec:panel}
\label{sec:m-telemetry}

\paragraph{Person-weeks} Our unit of analysis is the person-week (Monday to
Sunday): we relate an engineer's total token spend in a week to their total
commits in that week. Weeks 0 to 4 (2026-06-29 to 2026-08-02) form the
panel, and weeks $-7$ to $-1$ serve only as a pre-period. A person-week is in
the panel if it has any metered tokens or commits;
of the 2,030 person-weeks of the 406 metered engineers, we treat the 203
with neither as weeks the person was away and leave them out. The panel has
404 engineers (the other two used the agents only after week 4; 397 with
metered use in these weeks) and 1,827 person-weeks,
88\% with metered use, of which 12.9\% have no
commit. The median active week has 8 commits. The 178 engineers with no
metered use are not in the panel; in the pre-period
they committed slightly more than metered users (median 65 vs.\ 52 commits,
Mann--Whitney $p = 0.08$) and differ from them by at most 0.24 standard
deviations on any pre-period covariate. The gateway does not meter chatbots
or other assistants, so engineers outside the panel may use AI.

\paragraph{Analysis approach common to all RQs}
We answer each RQ between engineers and from week to week within an
engineer, since the two can disagree: heavy users may also ship a lot, yet
a busier week need not cost an engineer more than a slower one. We model
weekly token counts with Poisson regression estimated by quasi-maximum
likelihood~\cite{cameron2013count}, which keeps zero-token weeks and
estimates the mean consistently however dispersed spend
is~\cite{santossilva2006}. Within-person models include person and week
fixed effects~\cite{wooldridge2010}, i.e., they compare each engineer's
weeks with one another after removing the engineer's level and anything
common to all engineers in a week, such as a model release. A new frontier
model arrived in week 3 and accounted for 45\% of all spend within two
weeks; week effects absorb such a change only on average. For engineer
$i$ in week $t$, the within-person model is
\begin{equation}
  \label{eq:within}
  E[\text{tokens}_{it}] = \exp\!\left( c_i + d \log(\text{commits}_{it}) + t_t \right),
\end{equation}
where $c_i$ is the engineer's own level, $t_t$ the week, and the elasticity
$d$ is the percentage change in tokens per percentage change in commits.
Between-person slopes are least-squares fits on engineers' log averages
over the five weeks. Where we compare the two levels, we fit them jointly
in the within--between (Mundlak) form~\cite{mundlak1978,bell2015}, so that
their difference has a confidence interval. Confidence intervals are 95\%
percentile intervals from a cluster bootstrap over
engineers~\cite{cameron2015cluster}. The replication package
(Section~\ref{sec:availability}) gives the full specifications.

\paragraph{RQ1: How closely is an engineer’s token spend related to their output?}
\label{sec:m-coupling}
\label{sec:m-redundancy}
We answer RQ1 in two steps, which Section~\ref{sec:res-rq1} follows: how
closely a week's spend varies with that week's output, and what else in our
data explains the rest.

If spend rises in proportion to work, a week with twice the commits should
cost about twice as much, as cost-per-output metrics assume. We test this
with the elasticity $d$ in the Poisson model of Equation~\ref{eq:within}: $d
= 1$ means spend
doubles when commits double, and $d = 0$ means it does not vary with them.
The log of commits is undefined at zero, so we fit this model to the 1,347
engineer-weeks with both tokens and commits, from the 344 engineers who have
at least two such weeks to compare.
We also estimate the reverse slope, of commits on tokens; when spend and
output are closely correlated, the two slopes are near reciprocals. Two
things could make $d$ look smaller than it is. Commit counts only approximate a
week's work, which biases slopes toward zero~\cite{fuller1987}.
We therefore re-estimate $d$ by instrumental variables with lines changed as the
instrument~\cite{wooldridge2010}, which ignores squashing or splitting
commits because they change the count but not the lines. And output may
arrive later than
the spend behind it; we check this
with day-level cross-correlations of spend and commits at leads and lags of
up to ten working days. We repeat the estimate with lines changed and days
with a commit as the
output, at time units from one day to four weeks, with commit-count
categories in place of log commits, and with spend split into agent calls and tokens per
call, and test whether heavier-spend weeks' commits differ in size, platform
share, or issue linkage.

If spend does not co-vary with commits, it may reflect other parts of the work. We
first measure how much spend varies among weeks with similar output, as the
ratio of the 90th to the 10th percentile of weekly spend within each
quintile of weeks ranked by commits. We then ask how much of weekly tokens the
metrics in our data predict: commits, lines changed, bug-linked commits, and
days with a commit; then add issue-tracker activity, role, manager status, team
(grouped into divisions and product ecosystems), and each engineer's
\emph{prior work mix} (commit rate, commit size, and share of commits whose
message references a bug). For each set we compute the share of
deviance it explains relative to an intercept-only model, the Poisson
counterpart of $R^2$~\cite{cameron1996r2}, computed by
cross-validation~\cite{hastie2009} on held-out engineers (ten folds) for
differences between engineers and on held-out weeks (five folds) for changes
within an engineer. Finally, we
correlate each engineer's
spend with their teammates' average spend, excluding their own (a
leave-one-out mean), and repeat the within- and
between-person estimates on team-weeks.

\paragraph{RQ2: Where does the token spend that is not tied to repository output go?}
\label{sec:m-dark}
We answer RQ2 in two parts, which Section~\ref{sec:res-rq2} follows: from
the telemetry, how much spend is not tied to output and what the weeks with
such spend look like; and from the survey (Section~\ref{sec:survey-data}),
what engineers use the agents for and how many hours they report saving.

We cannot identify the spend not tied to output directly, because the
telemetry does not link agent calls to repository artifacts
(Section~\ref{sec:data}): in a
week with no output, none of that week's spend led to output; in a week with
output, we cannot tell which part did. We therefore estimate a
lower and an upper bound on the share, assuming that the spend and the
commits made with it happen in the same week. We test
this assumption with the lead--lag analysis of RQ1 and by repeating both
bounds at time units from one day to four weeks.

For the lower bound, the \emph{floor}, we count only the unambiguous case, the share
of all tokens spent in person-weeks with no commits,
restricted to weeks with above-median spend (other cutoffs in
Supplementary Material (SM)~\ref{app:results}). It misses spend not tied to
output in weeks with commits. For the upper bound, the \emph{curve estimate},
we predict an engineer's spend in a week with no output from how
their spend varies with output. It extends the Poisson model of
Equation~\ref{eq:within}, replacing $d\log(\text{commits})$ with
$b^\top s(\text{output})$:
\begin{equation}
  \label{eq:baseline}
  E[\text{tokens}_{it}] = \exp\!\left( c_i + b^\top s(\text{output}_{it}) + t_t \right),
\end{equation}
where $s(\cdot)$ transforms output to set the shape of the curve: raw counts,
square root, $\log(1+x)$, or a linear spline in $\log(1+x)$, each with
$s(0) = 0$. We sum the prediction for a week with no output,
$\exp(c_i + t_t)$, over person-weeks and divide by total spend to obtain
the share of tokens that does not vary with output. This share includes
the fixed cost of using an agent at all and the spend behind hand-written
output, so it overestimates.
Because the share depends on the shape, we fit all four shapes and compare
them on calibration in weeks with no output, out-of-sample deviance under
five-fold cross-validation~\cite{hastie2009}, and agreement with
negative-binomial and gamma models~\cite{cameron2013count}, which match the
Poisson mean only if the shape is right~\cite{santossilva2006}. We repeat
the estimate with commit size, issue type, and tool mix as predictors, and
within groups.

We also test whether spend precedes later output, by comparing an
engineer's commits, relative to their own mean, after a heavy-spend period
with no output with those after other periods. To relate a week's spend
beyond what its output predicts to how the agent was used, we correlate
each week's residual with the engineer's agent use that week (calls, days of
  use, context per call, share of generated text,
tool mix, and session length) and with the next week's bug-linked commits.
A session is a run of calls separated by at most 30 minutes.

\subsection{The survey}
\label{sec:survey-data}
\label{sec:m-survey}

\paragraph{The questionnaire} The company ran an anonymous internal survey
on the AI tools that employees use. Respondents indicated which
company-paid AI tools they had used in the past work week and were assigned
one of them to evaluate (the tool with the fewest evaluations at that
point); those who had used more than one could evaluate a second tool of
their choice. For each tool, respondents reported what they had used it for
in the past work week, what value it had provided, and whether it had
enabled work in the past month they would otherwise not have done. They
also reported whether it had saved or cost them time that week and how many
hours, how many colleagues the work would otherwise have involved, and its
effect on six aspects of their work on a five-point scale (productivity,
  quality of output, efficiency on routine and on complex tasks, cognitive
load, and finding information). A final question asked whether they had
handled colleagues' material they suspected was AI-generated and how much
time it cost them. Free-text questions asked for the main value, the work
enabled, and what saved and what cost time.

\paragraph{Respondents} The survey had 412 respondents (15.8\% of
employees; 525 evaluations in total): 400 evaluated at least one AI tool,
125 evaluated a second; 12 did not use any company-paid AI tool. Most
respondents (72\%) held
technical roles, and 16\% were team leads. The company invited employees
without incentives through an internal post that also asked colleagues
skeptical of AI to take part; 3\% of respondents indicated they did not
use any company-paid AI tool, a share consistent with the rest of the company.

\paragraph{What engineers use the agents for} For RQ2, we analyze the 197
evaluations of the two coding agents (99.8\% of metered tokens) by 184
respondents in technical roles. We report answers to the multiple-choice
questions as percentages of the evaluations that answered each question; a
respondent who evaluated both agents counts twice. \emph{Hours saved} is
the number respondents gave, negative when the tool cost time; it is
a self-report about the past week. We report medians and interquartile ranges.
The 289 answers to the four free-text questions were coded
thematically~\cite{braun2006} with an AI tool, and one author reviewed
every code manually; the codebook is inductive, per question, with
multiple codes per answer (1.6 on average). We report themes as shares of
substantive answers, with quotations.

\paragraph{Relating the answers} To see which kinds of use go with larger
reported time savings, we regress hours saved on the other answers at
once, because the answers overlap: respondents
who use the agent for more kinds of work also more often report enabled
work and name more colleagues that the work would otherwise have needed.
The predictors are coding use, the number of non-coding activity types,
enabled work, and the number of colleagues the work would otherwise have
needed; the controls are the tool, the division, and whether the week's
only reported value was experiments and learning. We model
hours on an inverse hyperbolic sine scale, which keeps zeros and losses,
standardize the coefficients, and compute bootstrap intervals over
respondents (SM~\ref{app:survey}). We also compare the coded free text with the
closed survey items.

\subsection{Ethical considerations}
This study examines workplace records and survey responses from employees of
JetBrains, where some of the authors work. Only one author had access to the raw
workplace records, and they pseudonymized them before sharing them. The
Carnegie Mellon authors had access to the pseudonymized data, in which
JetBrains employees are not individually identifiable, and used them only
for secondary analysis. The survey was anonymous and could not be linked to
individual telemetry records or to specific employees.
The data analyzed in this study are not used to evaluate individual employee
performance, and we report only aggregate results.
We do not release individual-level records or survey responses to protect
employee privacy. The study followed applicable ethical standards and the
ICC/ESOMAR International Code~\cite{icc_esomar_2025}.

\newpage
\section{Results}
\label{sec:results}


\subsection{RQ1: How closely is an engineer's token spend related to their output?}
\label{sec:res-rq1}

\subsubsection{A week's spend varies only weakly with that week's output}
\label{sec:res-coupling}

\begin{figure}[t]
  \centering
  \includegraphics[width=0.85\columnwidth]{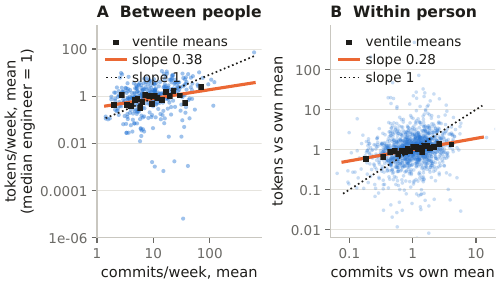}
  \caption{Spend vs.\ commits, both on log10 scales (slope 1 is
      proportional; lines are least-squares fits; squares are means of 5\%
    slices). (A) Engineers' average weeks. (B) Engineer-weeks relative to the
  engineer's own mean.}
  \label{fig:slopes}
\end{figure}

A week with twice an engineer's usual commits costs about 20\% more
tokens. The elasticity of tokens on commits, $d$ in
Eq.~\ref{eq:within}, is 0.26~[0.19, 0.32] (0.28 by least squares in
Figure~\ref{fig:slopes}B). In the other direction, the slope of commits on
tokens is 0.13~[0.03, 0.32], so a week with twice the usual tokens has about 10\% more
commits. If spend increased in proportion to output, both
slopes would be near one; their product is 0.03, and proportionality
($d = 1$) is rejected at every time unit ($p < 0.001$). Output itself varies
widely across weeks: an engineer's weekly commits typically differ from
their own average by a
factor of about two, i.e., half or double.

Two things could make $d$ look smaller than it is, but neither explains the
gap. First, commit counts are a noisy measure of work. Instrumenting commits with lines
changed raises the within-person slope $d$ to 0.48, still well short of
proportional. Second,
output may arrive after the spend: the day-level cross-correlation between
spend and commits peaks on the same day (0.33) and is gone after about four
working days (Figure~\ref{fig:lag} in SM~\ref{app:results}), so a week
contains most of the lag between spend and the commits made with it.
Nor does the result depend on how output and time are measured: the slope
is similar for days (0.30) and two-week periods (0.21), and with lines changed
(0.19) or days with a commit (0.40) as the
output. With commit-count categories and no shape imposed, spend rises along
the same constant-elasticity line at every output level
(SM~\ref{app:results}). Nor are the commits of heavier-spend weeks different
commits: no measure of their size, platform share, or issue linkage changes
at $p < 0.05$. A busier week brings more agent calls (0.25),
each with about the same number of tokens ($+0.01$, $p = 0.62$). The
within-person correlation is weakest among light users, 0.01 for engineers
who spend under a third of the median week compared with 0.37 for those
who spend over three times it, and every role, tool, and usage group stays
below proportional (SM~\ref{app:results}).

Between engineers, the association is somewhat stronger
(Figure~\ref{fig:slopes}A). On person means, the slope is 0.38 with each
engineer counted once and 0.44 with engineers weighted by their number of
weeks with metered use, compared with 0.26 within a person. In the other
direction, engineers who spend twice as much have about 8\% more commits
(slope 0.11). Fitted
jointly, the
between-minus-within difference is 0.15~[$-0.04$, 0.36], so engineers
who commit more may be heavier users beyond what their weekly output
explains, but five weeks per person is too short to be sure (Poisson fits
give larger slopes, driven by five engineers).

\subsubsection{Output and the other metrics explain little of the differences in spend}
\label{sec:res-redundancy}

\begin{figure}[t]
  \centering
  \includegraphics[width=0.85\columnwidth]{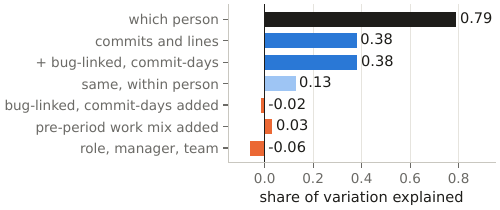}
  \caption{Share of variation in weekly token use that each set of
    predictors explains. ``Which person'' is fit in sample; ``same, within
    person'' is tested on held-out weeks; every other set on held-out
    people. ``Added'' bars are separate fits. A negative share means the set
    predicts held-out people worse
  than the week's average spend alone.}
  \label{fig:explained}
\end{figure}

Output alone explains little of the spread in spend. Within every quintile of
weekly commit counts, the engineer-week at the 90th
percentile of spend uses 39 to 240 times the tokens of the one at the 10th
(Figure~\ref{fig:spread}).

The four activity metrics (commits, lines changed, bug-linked commits,
commit-days) explain 38\% of the between-person
variation in weekly tokens on held-out people and 13\% of the within-person
variation on held-out weeks (Figure~\ref{fig:explained}). Commits and
lines changed alone do just as well (38\% between); bug-linked commits and
commit-days add nothing ($-2$ points), nor does issue-tracker
activity, and how engineers
worked in the pre-period (their commit rate, commit size, and share of
bug-linked commits) adds 3 points. Given all four, the 90th-percentile
person-week still uses 58 times the 10th.

The engineer's role and team explain even less. Role, the manager flag, and
team together predict held-out people worse than each week's average spend
does ($-6\%$), possibly because spend varies widely among engineers on
the same team: an engineer's spend
correlates at 0.24 with the leave-one-out mean of their division or
ecosystem and at 0.13 with that of their team. Summed to teams, spend is no more closely
related to output than for individuals: the within-team slope is 0.31 for
team paths and 0.27 for product ecosystems, compared with 0.25 for the same
engineers (SM~\ref{app:results}). The person does account for the spread,
explaining 79\% of the deviance in our sample.

Which engineers spend the most is stable over the five weeks; who has the
highest cost per commit is not (SM~\ref{app:results}).
\label{sec:res-stability}

\subsection{RQ2: Where does the token spend that is not tied to repository
output go?}
\label{sec:res-rq2}

\subsubsection{Up to two fifths of spend is not tied to the week's output}
\label{sec:res-dark}

Weeks with no commit account for 4\% of all tokens, but by the curve estimate
the share that does not vary with the week's output is about two fifths. Most
of the spend not tied to output falls inside weeks that did
produce commits.
Throughout this section, ``output'' means commits. Time spent on code
review, design, or repositories outside our sample therefore counts as not
directly tied to output, as does time spent on exploration, learning, and
abandoned attempts.

\paragraph{The floor} The lower bound is small. Of the 1,599 engineer-weeks
with any metered use, 207 have no commit, mostly light-use
weeks. The 49 with above-median spend
account for 3.0\%~[1.8\%, 4.5\%] of all tokens; all 207
together account for 4\%. By construction, the floor depends on the time
unit; it is 25\%
for single days and 1\% for two-week periods. We
use the week because spend and its commits usually fall within four
working days (Section~\ref{sec:res-rq1}).

\begin{figure}[t]
  \centering
  \includegraphics[width=0.85\columnwidth]{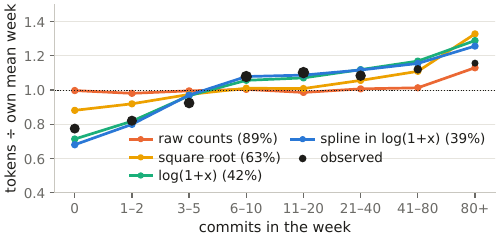}
  \caption{Tokens relative to the engineer's average week, by the
    week's commit count: observed (dots) and predicted by each curve shape
    (lines), with the share of tokens attributed to zero output under each
  shape in parentheses.}
  \label{fig:dark}
\end{figure}

\paragraph{The curve estimate} The upper bound is an order of magnitude
larger. A week with no commit still costs 77\% of the engineer's average
week (Figure~\ref{fig:dark}), and the share of all tokens this implies
depends on the shape of the curve. Raw counts attribute 89\%~[79\%, 97\%] of all tokens
to zero output, the square root 63\%, $\log(1+x)$ 42\%~[33\%, 51\%],
and the spline 39\%~[31\%, 46\%]. We trust the
last two. Spend rises steeply over the first few commits of a week and then
flattens, a shape an exponential-in-counts curve cannot take, and all three
diagnostics from Section~\ref{sec:methods} rank counts last
(SM~\ref{app:results}). Adding commit size and tool mix to the model, or
fitting it to the first four weeks alone, changes the estimate by at most a
few points (SM~\ref{app:results}). Even so, 39\% remains an upper bound, since
it includes the fixed cost of using an agent at all and the spend behind
commits written by hand (Section~\ref{sec:methods}).

\begin{figure}[t]
  \centering
  \includegraphics[width=0.85\columnwidth]{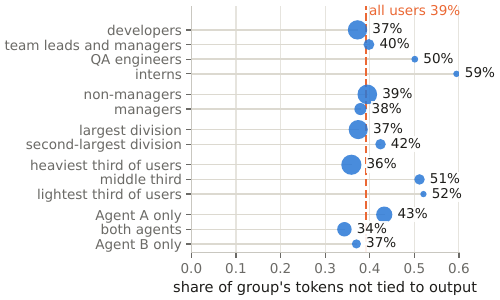}
  \caption{Share of each group's tokens not tied to output, by the spline.
    Dot area is the group's share of all tokens; the dashed line
  marks all users.}
  \label{fig:darkgroups}
\end{figure}

\paragraph{Where the share differs} We split the share by role and division,
and by how much engineers use the agents (Figure~\ref{fig:darkgroups}).
Across the groups that
spend most of the tokens, the share barely varies: developers, team leads and
managers (a
role inferred from job titles), and each of the two large divisions range
from 37\% to 42\%. It is higher only where the job is expected to produce
fewer commits or
the agent is used less: 50\% for QA engineers, 59\% for interns, and 51\% and 52\%
for the middle and lightest thirds of users, relative to 36\% for the
heaviest third, who
spend 79\% of all tokens. The share varies far more between individual
engineers than between any of these groups: per engineer the median is
46\%, the middle 80\% of engineers range from 31\% to 68\%, and the 10th
and 90th percentiles of each engineer's level of such spend differ 59-fold.
Team, role, the
manager flag, and prior work mix together explain only 11\% of this
variation on held-out people, as they explained little of spend overall
(Section~\ref{sec:res-rq1}).

\paragraph{What the weeks with excess spend look like}
A week's spend beyond what its commit output predicts could reflect a
costlier model or agent, or the same tools used differently. Our data point
to both. In such weeks engineers used more of the newly released model:
within person, the residual correlates with the week's share of tokens on
that model at 0.15 ($p < 0.001$), though not with the share from either
agent or from frontier models as a group, which includes older frontier
models. They also worked in longer sessions: 31\% of calls fall in sessions
longer than two hours, compared to 18\% in the person's other weeks.
Generated text is under 1\% of tokens in either kind of week (0.6\% vs.\
0.9\%), so the excess is context sent to the agent rather than code
returned. The excess also does not predict bug fixes
in the following week: the residual correlates with that week's bug-linked
commits at 0.03.

\subsubsection{Engineers use the agents for far more than coding, and report
saving more hours the broader that use}
\label{sec:res-uses}

\begin{figure}[t]
  \centering
  \includegraphics[width=0.85\columnwidth]{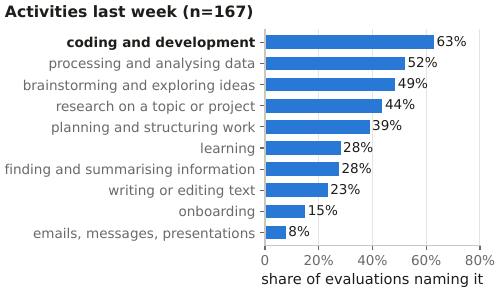}
  \caption{Activities for which technical respondents reported using the two
  coding agents in the past work week. ($n = 167$; multiple selections allowed).}
  \label{fig:activities}
\end{figure}

\begin{figure}[t]
  \centering
  \includegraphics[width=0.85\columnwidth]{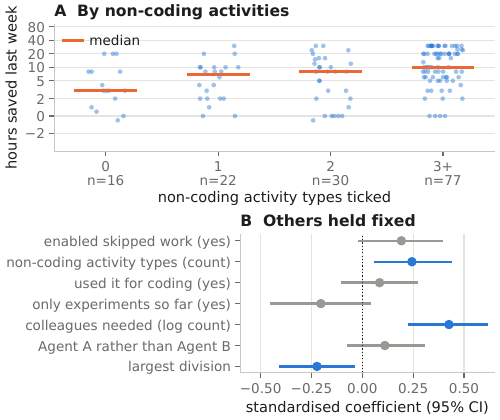}
  \caption{Self-reported time savings or losses from coding-agent use among
    technical respondents. (A) Hours saved or lost in the past work week,
    grouped by the number of non-coding activity types reported. Each dot is
    one evaluation and
    the bar its group's median; the vertical axis is compressed above 2 hours.
    (B) Associations with reported time savings from a regression model
    including all seven predictors simultaneously (134 evaluations; R² = 0.43).
    Predictors were standardized, and the outcome was asinh-transformed. Points
    show coefficient estimates; lines show 95\% bootstrap confidence intervals.
  Blue intervals exclude zero.}
  \label{fig:survey-model}
\end{figure}

Of the
197 evaluations of the two coding agents by technical respondents, 167
answered the questions about past-week use. In 91\% of these, respondents
reported that the tool had saved them time that week (median 10 hours,
interquartile range 5--20), and in 88\%, they reported a positive effect
on overall productivity. Respondents reported using these agents for
coding in 63\% of evaluations and for at least one non-coding activity in
89\% overall (83\% among the 100 evaluations of assigned tools): processing
and analyzing data (52\%), brainstorming
(49\%), research on a topic (44\%), planning work (39\%), learning (28\%),
and writing (23\%) (Figure~\ref{fig:activities}). Median hours saved rise
from 3 for respondents who used the agent only for coding to 10 for those
who reported two or more non-coding activities
(Figure~\ref{fig:survey-model}A). In the model of
Section~\ref{sec:m-survey}, which relates hours saved to all of these
answers at once (Figure~\ref{fig:survey-model}B), the count of non-coding
activity types is associated with hours saved (0.24~[0.06, 0.43]), as is
the number of colleagues otherwise needed (0.43~[0.23, 0.61]), but coding is
not (0.08~[$-0.10$, 0.27]). Enabled work
has a similar estimate whose interval includes zero (0.19~[$-0.01$,
0.39]). Respondents who report enabled work also chose more activities and name
more colleagues, so with 134 answers we cannot separate the three
(SM~\ref{app:results}).

\begin{figure*}[t]
  \centering
  \includegraphics[width=\textwidth]{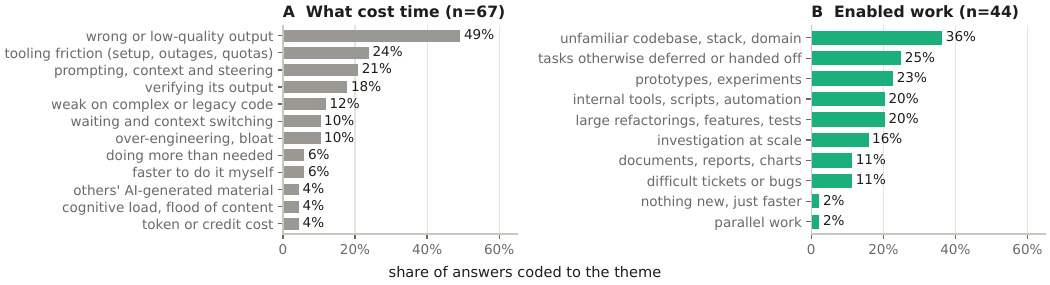}
  \caption{Thematic coding of the free-text answers on what cost time and on
    enabled work.
    Figure~\ref{fig:themes-rest} in
  SM~\ref{app:survey} shows the other two questions.}
  \label{fig:themes}
\end{figure*}

\begin{figure*}[t]
  \centering
  \includegraphics[width=\textwidth]{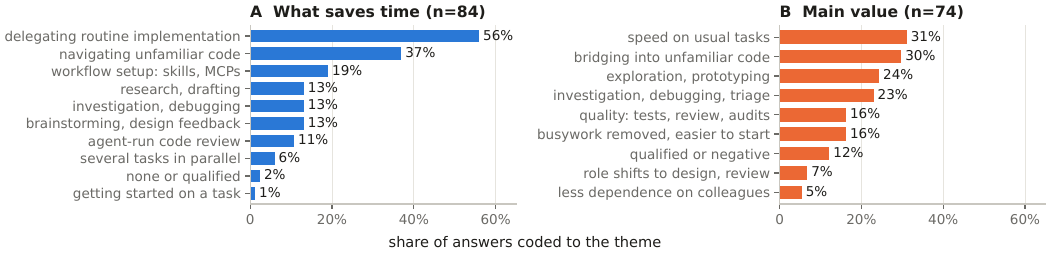}
  \caption{Thematic coding of answers by respondents in technical roles to
    open-ended questions about what saved them time while working with agents
    and the agent’s main perceived value (AI-assisted coding, checked by one
  author), as in Figure~\ref{fig:themes}.}
  \label{fig:themes-rest}
\end{figure*}

In their own words, respondents describe much of the coding-agent use as
work outside their usual tasks: in unfamiliar code, domains, or
technologies, on tasks they would otherwise have deferred, and in
exploration. Asked for the agent's main value, they split between speed
on their usual tasks (31\% of 74 substantive answers) and work in
unfamiliar code, domains, or technologies (30\%); 24\% described
exploration and a wider set of tasks worth attempting (``I can try things
  that I was scared to try''; Figure~\ref{fig:themes-rest}B in
SM~\ref{app:survey}). The 44 substantive descriptions of work done in
the past month only because the agent existed show the same pattern. Of these,
36\% describe work in an unfamiliar codebase, language, stack, or area
(``fixed a bug in [system] despite having no prior knowledge of that codebase'').
Another 25\% describe tasks that would otherwise have been deferred or
handed to someone else (a type-inference problem ``that would otherwise take
weeks so we would never have picked it up''). Prototypes and experiments
account for 23\%, internal tools 20\%, and documents and reports
11\% (Figure~\ref{fig:themes}B). Only one of the 44 says the work would have
happened anyway, just slower. Much of this work produces understanding
before it produces a commit, or no commit at all in our data.
Asked what kind of value the week's tasks created, 59\% named ``experiments
and learning that have not yet brought tangible value''.

\begin{figure}[t]
  \centering
  \includegraphics[width=0.85\columnwidth]{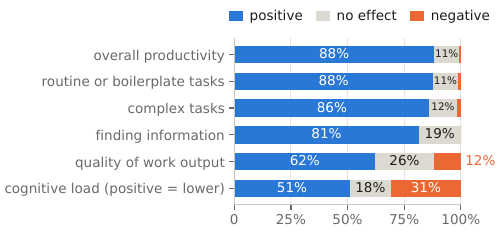}
  \caption{Effect of the coding agents on six work aspects,
  tech.\ respondents.}
  \label{fig:effects}
\end{figure}

Respondents see part of agent use as not producing valuable output. Of the 67
substantive answers on what cost time, 49\% describe wrong,
plausible-but-wrong, or low-quality output (``a
plausibly-looking but ultimately [useless] symptom fix''). Another 24\%
describe tooling friction unrelated to the model (setup, credentials,
outages, quota limits), 21\% the cost of prompting, context management, and
steering, and 18\% the cost of verifying output, which sometimes exceeds the
time saved (``there is a point where time spent on prompting and reviewing
exceeds the time savings''; Figure~\ref{fig:themes}A). Naming a cost is
not associated with a lower overall productivity rating. Respondents whose answer was
coded as non-valuable
output rate productivity at 4.2 compared with 4.3 and report the same median
hours saved (9). The survey's hours question asks only how much
time the respondent's own tool saved or cost them, so two costs that other
questions measure are not part of that figure: the cognitive load of
working with agents, and the time spent on colleagues' AI-generated
material.
Among technical agent users, 31\% report higher cognitive load because of AI
agents and 51\%
lower (Figure~\ref{fig:effects}). The cognitive-load rating is nearly
uncorrelated with reported hours saved ($\rho = 0.11$, compared with 0.57
for overall productivity); respondents describe the trade (``I do more, but my brain
gets much more tired''). Colleagues bear part of the cost: 55\%~[49\%,
60\%] of the 299 who answered report handling colleagues'
material they suspected was AI-generated, costing a mean 1.6
hours a week~[1.2, 2.1].
Team leads handle it more often (73\% vs.\ 50\%; odds ratio 2.7,
$p = 0.006$), with no more reported benefit.

\section{Discussion}
\label{sec:discussion}

We discuss what token spend measures, what that implies for its three uses
(measuring productivity, monitoring adoption, and setting budgets), and what
to measure instead.

\smallskip\noindent\textbf{What token spend measures.}\label{sec:disc-measures}
\looseness=-1
Token spend records how much an engineer uses agents, scaled by the context
included in each call and by the model in use, and in our data it varied far
more between engineers than with the week's
output. Who the engineer was accounted for 79\% of the variation in weekly tokens in
our sample; on held-out data, output metrics predicted under two fifths of
the differences between engineers and 13\% of the variation within them,
and role and team alone predicted held-out engineers worse than the weekly average. Within an engineer, a
week with twice
the usual commits had only about a fifth more tokens
(Section~\ref{sec:res-rq1}). Token spend has been called ``the lines-of-code
metric of the AI
era''~\cite{bilicarar2026}, and our data support the comparison: both are
easy to count and loosely related to what engineers accomplish. The two
differ in what they count. Lines of code count output, and stayed in use
because program size sufficiently reflected the effort behind it
(Section~\ref{sec:rw-metrics}); tokens count an input, recorded whether or
not anything is produced, and in our data they reflect output only
weakly.
Organizations should therefore treat
spend as a record of agent use and not as a measure of output.

Prior estimates and ours agree that output rises far less than spend. Ten
times the spend corresponds to about 40\% more output across Jellyfish's
usage deciles~\cite{jellyfish2026tokenmaxxing}, to twice the output in a
press summary of vendor data~\cite{fernholz2026tokenmaxxing}, and to at most
76\% more under the mandate He et al.\ study~\cite{he2026} (our calculations,
assuming constant elasticity). In our data (the slope of commits on tokens) it corresponds to about 30\% more output between engineers and 35\% within [7\%, 109\%]. The prior figures compare
different engineers,
grouped by spend, or follow an organization over time, during which other
things changed. Our within-engineer estimate holds the engineer fixed and is
similar, so the weak association is not only an artifact of comparing
different engineers. Engineers chose their weekly spend, so this estimate is still not the effect that an imposed budget or target would have.

By our best-fitting curves, up to about two fifths of spend does not vary with the week's commits. This upper
bound includes the fixed cost of using an agent at all and the spend behind
commits written by hand (Section~\ref{sec:res-dark}); the rest may be work
the repository does not record, or waste. In the survey, respondents reported uses that need not end in a commit
(analysis, research, planning) and named wrong or low-quality output as the
most common cost (49\% of answers; Section~\ref{sec:res-uses}), so agent use
in general includes both.  A token spent on a
prototype that informs a design
counts the same as one spent on output that is discarded, so the count
alone cannot separate the two.

The setting may limit how far our estimates transfer. JetBrains builds AI
coding tools, so some metered use may be testing of its own agents; its
engineers work on established codebases, the type where He et al.\ report smaller AI
gains~\cite{he2026}; and use was voluntary, with no target or mandate, so
differences in spend reflect, in part, engineers' own choices.

\smallskip\noindent\textbf{Measuring productivity.}\label{sec:disc-performance}
Leaderboards~\cite{fortune2026meta}, spending expected in proportion to
salary~\cite{nieva2026forbes}, and minimum spending
targets~\cite{orosz2026tokenmaxxing} all reward higher spend, and the salary comparison equates it with better
engineering. Some companies have since dropped such
measures after reports
that engineers inflated their usage to rank higher: Meta told engineers
that token counts will not be used to evaluate their
impact~\cite{cooter2026meta}, and Amazon shut down a usage leaderboard and
called token usage a ``deeply flawed'' measure of teams' AI
use~\cite{palmer2026amazon}. Gaming is the familiar fate of a measure that
becomes a target~\cite{strathern1997improving}, but our data suggest that
spend is a weak measure of output even without a
target: in our setting, where the company did not set a spending target, the 90th
percentile of weekly spend within each commit quintile was 39 to 240 times
the 10th. If the within-engineer association were
causal, doubling an engineer's spend through a target would raise their commits by
about 10\% (2\% to 25\%). Removing a leaderboard addresses the gaming but not
these weaknesses,
so companies should not evaluate engineers' output or impact by their spend,
even where no one games it, and should not expect spending targets to raise output much.

Dividing spend by output, as analytics vendors do when they report token cost
per merged pull request as a return on
investment~\cite{jellyfish2026tokenmaxxing,weave2026index}, does not show the return on the spend, or how efficiently engineers use the agent. First, the ratio falls, on average, as
commits rise: a week with twice an engineer's usual commits has about 1.2
times the tokens, so its tokens per commit are $1.2/2 = 0.6$ times the
usual, i.e., 40\% lower. An engineer thus tends to look more efficient in a
week with more commits. Second, the ratio ranks engineers consistently overall (projected
reliability 0.80) but not at the top: only 17 of the 34 engineers in the top
decile by tokens per commit in odd weeks remain there in even weeks
(SM~\ref{app:results}). Third, a company that divides tokens by commits
treats every token as going to a commit and every commit as coming from
tokens, so agent work that ends without a commit counts as cost without
return, and a commit written by hand as return without cost. Fourth, the
ratio changes with the model: in our data, weeks on the newer model had about
45\% more tokens at the same commit count
(SM~\ref{app:results}), so a company's ratio can
rise with a model release without any change in commits. Companies that want
to judge whether engineers use agents efficiently
should compare their spend on comparable tasks, which requires the task
records described below.

\smallskip\noindent\textbf{Monitoring adoption.}\label{sec:disc-adoption}
A token-based count of adopters depends on the token threshold chosen to
define an adopter. When Steve Yegge called AI adoption at Google shallow,
Google answered with 40,000 weekly agent users and Yegge with a threshold of
four million tokens a day~\cite{bi2026yegge,venturebeat2026,yegge2026}. Both
numbers can be right, because they count the two ends of one skewed
distribution. In our data, the heaviest third of users accounted for 79\% of
all tokens, so a headcount of weekly users can be high
while few engineers cross a high threshold. Because weeks on the newer model
had more tokens at the same commit count, the share above a fixed threshold
would also likely rise with a model release. A token threshold is arbitrary, and the share above it changes with model
releases, so organizations that monitor adoption could report the weekly headcount
alongside a count that does not scale with context size, such as the share
of engineers who use agents on most working days.

\smallskip\noindent\textbf{Setting budgets.}\label{sec:disc-cost}
To decide how much to spend, companies need to know what the spend returns.
One vendor reports merged output per dollar as a ``return on token spend''
and finds that it fell by about 70\% during 2026~\cite{weave2026index}. Such a decline need not mean the added spend was wasted, for two reasons. When output rises
less than proportionally with spend, as it did with tokens within engineers
in our data, output per unit of spend declines as spend grows even if the
additional spend serves useful work; and output per dollar changes with
token prices and models. In output per dollar, only merged work counts as
output, so companies that
use it should weigh benefits that commits do not capture.
Respondents report several, which self-reports can overstate~\cite{becker2025metr}: hours saved, tasks they would otherwise have
skipped, and work that would otherwise have needed colleagues
(Section~\ref{sec:res-uses}). The value of these benefits depends on how the
work is used later, for example whether a prototype informs a decision, and
need not grow with the tokens spent on it. Companies should weigh these benefits against costs that are not on the
bill: 31\% of technical agent users
reported higher cognitive load from agents, and 55\% of the 299 respondents who answered reported handling colleagues'
material they suspected was AI-generated, from any AI tool, at a mean cost
of 1.6 hours a week.

For setting budgets, spend is more useful summed over several weeks:
five-week totals rank engineers consistently (projected rank reliability 0.87), whereas two single weeks rank-correlate at only 0.4
to 0.6. With ranks this stable, a team's past spend shows which
members will need the most budget; the budget's level should be revised
after a model release. Cutting spend is harder. A company may cap it per
person~\cite{constantin2026tokenminimizing}, which limits mostly the
heaviest users. Their share of spend not tied to output is
the lowest (36\%, vs.\ 51\% and 52\% for the other thirds), plausibly because the fixed cost of using an agent weighs less in their spend; but because they
spend 79\% of all tokens, they also account for most of the spend not tied
to output (Section~\ref{sec:res-dark}). A company that caps therefore cuts
spend tied to output along with spend
that is not, and the token count does not show which. Lowering what
each call costs may avoid that problem, since it leaves engineers' use of
agents unchanged. Most tokens are context sent to the agent, much of it re-sent on every turn, and in weeks with more spend than their output predicted,
engineers used more of the newly released model and worked in longer
sessions, in which that context accumulates. Companies that need to cut could therefore try managing or
compressing context~\cite{gao2026turn,xiao2026trajectory,sun2025tokensugar},
caching repeated context, and routing routine calls to cheaper models before
capping engineers. They should measure billed cost rather than tokens, since
reducing context can raise cost when the agent re-reads what was
removed~\cite{weinberger2026token}.

\smallskip\noindent\textbf{What to measure instead.}\label{sec:disc-measure}
Organizations want to know what agent-assisted engineering is worth; Amazon
now asks its teams to measure long-term outcomes~\cite{palmer2026amazon}.
Billing records supply the inference cost and the token count supplies the amount
of agent use; the value needs other records. Two changes would begin to
provide them, although some of
that value, such as what engineers learn, may remain hard to measure. First,
record at the gateway which issue or task each session served. With
such a record, a company could join spend to each task's outcome, such as
a substantial product improvement or a decision that a prototype
informed, rather than only to whether it produced code. Second, develop
measures of the value of agent-assisted work. SPACE and
DORA~\cite{forsgren2021,forsgren2018accelerate} were designed for code that
people wrote, and their metrics become noisier when the same number can come
from different mixes of human and agent work~\cite{houck2026}. We know of no validated measures of value across the outer loop of
development, from what ships to how products and businesses improve; until there are, a company can record how often engineers report taking on
tasks they would otherwise have skipped (58\% of coding-agent evaluations
here; SM~\ref{app:results}), as a sign of new work rather than of its
value.

Research that measures the outcomes of AI use through commits and pull
requests~\cite{agarwal2026ides,he2026} observes only the agent work that
ends in a commit; joining those records to agent telemetry would bound how
much agent work
does not. Analyses
that use spend as exposure, such as comparisons of usage
deciles~\cite{jellyfish2026tokenmaxxing}, should separate variation within
engineers from variation between them, because spend is largely a stable trait of the engineer
(Section~\ref{sec:res-stability}).



\section{Conclusion}
\label{sec:conclusion}

In this paper, we asked what an engineer's token spend measures, and whether it is, as one critic has claimed~\cite{bilicarar2026}, the lines-of-code metric of the AI era. In
five weeks of metered agent use and commits from 404 engineers at one
company, spend varied far more between engineers than with their output: a
week with twice an engineer's usual commits had only about a fifth more
tokens, and up to two fifths of spend did not vary with the week's commits.
Survey respondents reported using agents for analysis, research, and
planning as well as for coding. Token spend therefore records how much
engineers use agents, and, like lines of code, it says little about what
that use produced. Because spend was weakly related to output even where the company had not
set a spending target, we recommend that companies not evaluate engineers'
output by their spend; and because output rose far less than spend, that
they not expect spending targets to raise output much, or divide spend by
commit or pull request counts to judge how efficiently engineers use
agents. Two steps would begin to make the value of agent-assisted work measurable:
records that link agent sessions to tasks, and measures of the value of work
that people and agents do together. We hope our results motivate research on both.

\section{Data availability}
\label{sec:availability}

The raw data and survey responses cannot be published, but anonymized records
may be available under a non-disclosure agreement; interested readers should
contact the JetBrains authors.
Our replication package
\href{https://github.com/CMUSTRUDEL/token-study-replication}{github.com/CMUSTRUDEL/token-study-replication}
contains the analysis scripts and supplementary material.

\smallskip\noindent\emph{Acknowledgment.} We used AI
for help with data analysis and reporting. All research design decisions
are our own; we reviewed all outputs and take full responsibility.

\balance
\bibliographystyle{IEEEtran}
\bibliography{references-reframed}

\clearpage
\appendices
\renewcommand{\appendixname}{SM}

\section{Additional results}
\label{app:results}

This section of the supplementary material reports further results,
sensitivity analyses, and subgroup
estimates for Section~\ref{sec:results}, in the order of the RQs.

\begin{figure}[t]
  \centering
  \includegraphics[width=\columnwidth]{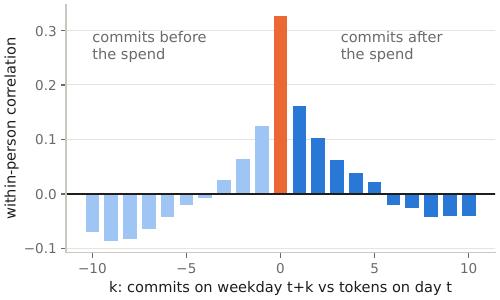}
  \caption{Within-person correlation between an engineer's tokens on a working
    day and their commits $k$ working days later, both as $\log(1+x)$ with
    person and day means removed. Small negative values at long lags are
  expected once each engineer's mean over five weeks is removed.}
  \label{fig:lag}
\end{figure}

\paragraph{Coupling (RQ1)} With commit-count categories and no shape assumed, spend
rises along the constant-elasticity line, and the day-level cross-correlation
is slightly stronger one to three days forward than backward (0.16, 0.10,
0.06 vs.\ 0.13, 0.06, 0.03; Figure~\ref{fig:lag}). Split by average
weekly commits (under 5, 5--12, 12--25, over 25), the within-person slopes
are 0.07, 0.26, 0.28, and 0.27. Split by average weekly tokens, the
within-person correlation is 0.01 for engineers under about 0.3 times the
median week, 0.08 for 0.3 to 1 times, 0.25 for 1 to 3 times, and 0.37 for
those over 3 times. By dominant tool, the slope is 0.20 on Agent A-dominant
weeks and 0.49 on Agent B-dominant weeks (65 people), and by job family it
ranges from 0.22 for developers to 0.30 for team leads, 0.41 for interns,
and 0.62 for QA engineers (differences significant at $p = 0.04$ after
correction); every group is far below one.

\paragraph{Teams (RQ1)} Summing the panel to team-weeks does not make spend
more closely related to output. We estimate the within-team slope of
tokens on commits, with team and week effects, for the 28 team paths with
at least five metered members (176 engineers, median 6 per team): it is
0.31 (95\% team-cluster bootstrap CI [0.16, 0.50]), and 0.28 with the
number of active members as a control. For the 22 product ecosystems with
at least five members (376 engineers), it is 0.27~[0.11, 0.41], and
0.21 with the control. On
the same engineers, the within-person slope is 0.25. Summing any group of
engineers removes some week-to-week noise, so we compare each estimate
with random groups of the same sizes drawn from the same engineers: they give
0.27 for teams and 0.21 for ecosystems, and 32\% and 26\% of random
groupings reach the actual slope, so the team structure adds nothing we can
detect. Least squares on logs gives higher within-team slopes (0.37 and
0.46), which only 8\% and 7\% of random groupings reach, so aggregation
may strengthen the association on the log scale; under the Poisson
model used throughout, it does not. Between teams, per active member, the
slope is 0.10~[$-0.28$, 0.56] for teams and 0.29~[$-0.21$,
0.80] for ecosystems.

\begin{figure*}[t]
  \centering
  \includegraphics[width=\textwidth]{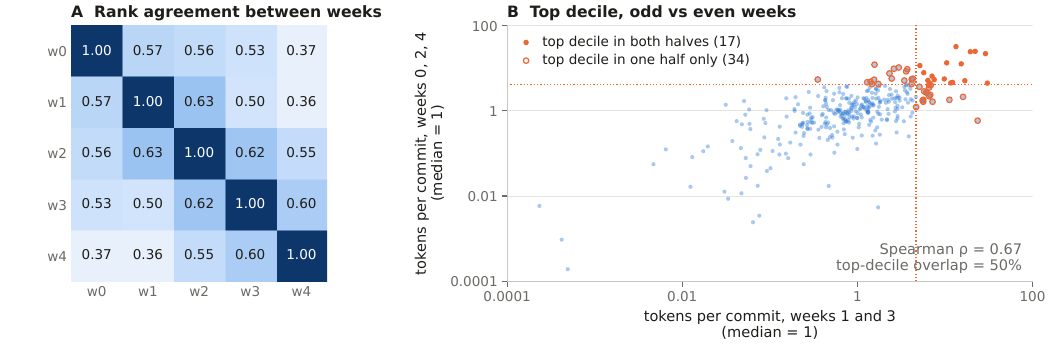}
  \caption{(A) Rank correlation of weekly tokens between every pair of
    analysis weeks, over engineers active in both. (B) Tokens per commit over
    weeks 0, 2, and 4 vs.\ weeks 1 and 3; filled dots are in the top decile
    of both halves (17 engineers), open dots in the top decile of only one
  (34 engineers).}
  \label{fig:stability}
\end{figure*}

\paragraph{Stability (RQ1)} If spend is a property of the engineer, an
engineer's rank should persist across weeks. For spend, commits, the
engineer's level of spend given output ($c_i$ at the common elasticity),
and tokens per commit, we split each engineer's weeks into odd and even
and compute the Spearman rank correlation between the two halves,
projected to five weeks with the
Spearman--Brown formula~\cite{spearman1910,brown1910}. The split-half
correlation is 0.77 for
spend and 0.77 for commits, which the formula projects to 0.87 for the full
five-week total; it is 0.84 for the level and 0.80 for tokens per commit
(split-half 0.72 and 0.67; Figure~\ref{fig:stability}B).
Adjacent weeks rank-correlate at only about 0.6 and weeks four apart at
about 0.4 (Figure~\ref{fig:stability}A). Output is stable too: an
engineer's commit rates in the pre-period and in the analysis weeks
correlate at 0.70, and spend in the analysis weeks is almost as closely
related to engineers' pre-period output ($\rho = 0.32$) as to their
concurrent output ($\rho = 0.44$). A
per-person elasticity would need far more productive weeks than the median
of four here; computed anyway, it correlates across halves at
$\rho = -0.07$. Cost per commit is stable overall but not at the top. Because
spend rises
much more slowly than output within a person ($T/Y \propto Y^{d-1}$),
weeks with more commits have lower tokens per commit (Spearman $\rho =
-0.51$ within a person, $-0.32$ between engineers), and of the 34 engineers
in the top decile by tokens per commit in odd weeks, only 17 remain there in
even weeks (Figure~\ref{fig:stability}B).

\paragraph{The floor and the curve (RQ2)} With the person's own median as the
cut, the floor is 2.4\%. Counted by three-day blocks it is 10\%, and by
four-week blocks 0.4\%. At shorter time units, the ordinary one-day delay
between spend and its commit appears as spend not tied to output; at longer
ones, a week without commits is hidden behind a single commit. For a week
with no commit, the square-root and $\log(1+x)$
forms predict 88\% and 71\% of the engineer's average week, the raw-counts form
100\%, and the spline 68\%, compared with the observed 77\%; the spline is within a
few percent of the observed values
in every other bin. We measure calibration as observed over fitted tokens,
summed over weeks with no output (so weighted by tokens, unlike the
per-engineer 77\%). It is 0.62 under counts, 0.71 under the square root, 0.91 under
$\log(1+x)$, and 0.96 under the spline, so the counts form predicts far more
spend for those weeks than they held. Under counts, the Poisson,
negative-binomial, and gamma shares are 84\%, 60\%, and 64\% on our output
set, 24 points apart, a sign that the mean is misspecified; under
$\log(1+x)$ they are 42\%, 39\%, and 37\%. In five-fold cross-validation the
$\log(1+x)$ and spline forms have the lowest held-out deviance for every
output set, and counts the highest. With a natural cubic spline in raw
commits, with one interior knot at the median positive-commit week as in
the linear spline, the estimate is 45\% (calibration 0.86). Adding lines
changed to the count
leaves the spline share at 38.9\%, and adding the type of issue a commit
names (bug, task, feature) moves it by under a point. By
day the spline estimate is 77\%, by three days 53\%, and by
two weeks 40\% ($\log(1+x)$ on our output set: 73\% by day and 48\% by
two weeks); as with the floor, the day-level figure is inflated because two
thirds of days have no commit. With size and work type in the model,
adjusting for the share of tokens from each agent and from the new model
moves the $\log(1+x)$ estimate from 44\% to 45\%, since tool mix is not
output. Tool mix is still associated with consumption: a week with all its
tokens on the new frontier model has 45\% more tokens
than one with none, at the same commit count ($p = 0.03$). The first four
weeks alone give 46\%.

\paragraph{Groups and trend (RQ2)} By dominant tool, the share is 43\% for
Agent A-only engineers, 37\% for Agent B-only, and 34\% for those using both.
The top tenth of people by this component account for 38\% of it compared
with 45\% of all tokens, so heavy users' tokens are more tied to their output than light
users'. Refitted on the first two analysis weeks alone, the share is 39\%, and
on the last three 30\%, while week-by-week fits of the $\log(1+x)$ form stay
between 42\% and 44\% with no trend. Five weeks is too short to say whether the
difference is a trend, and the trajectory with experience cannot be
identified from a panel in which 162 people already have token use on its
first day.

\paragraph{What the spend precedes (RQ2)} At the day level, a working day
with above-median token spend and no commit is followed by 1.18 times the
person's average daily commits, compared with 1.20 after a day with commits, 1.00
after a no-commit day with light spend, and 0.62 after a day with no
activity; the difference from the light no-commit day is $+15\%$ within
person ($p = 0.24$). At the week level, heavy
no-commit weeks are followed by 150\% more commits than light ones ($p =
0.008$, 148 weeks); among weeks with four or more active days the estimate is
similar but imprecise ($+146\%$, $p = 0.08$, 65 weeks), so the records cannot
separate a catch-up in commits from the light group returning from absence.
High-residual
weeks have 1.8 times the calls per active day, 1.6 times the calls per
session, 1.4 times the tokens per call, and a lower share of generated text
($-0.52$; 0.6\% vs.\ 0.9\%); within person, the residual correlates with
agent calls at 0.74, with days of use at 0.59, and with context per call at
0.53. The shares of tokens from either agent and from frontier models as a group
do not distinguish them; the share on the newly released model does
(within-person correlation 0.15, $p < 0.001$). The share of
commits naming a bug is 12.0\% after high-residual weeks compared with 13.0\% after
the person's other weeks, and next week's bug-linked commits change by $-2\%$ per
unit of residual with next week's total commits held fixed ($p = 0.75$),
whereas a friction account predicts an increase. Technical survey respondents likewise
report a median of one hour a week reading colleagues' AI-generated material
and none reworking it.

\paragraph{Enabled work in the survey model (RQ2)} On the closed enablement
question, respondents in 58\% of the technical coding-agent evaluations
said the agent let them complete tasks in the past month they would
otherwise have skipped.
These respondents say they would otherwise have needed two colleagues rather
than one, and report a median of 12 hours saved in the past week compared
with 4 for the rest and a productivity rating of 4.5 compared with 4.1. With activity
breadth and colleagues named held fixed (Figure~\ref{fig:survey-model}B), the
hours gap is no longer separable from them (0.19, $-0.01$ to 0.39):
respondents reporting enabled work also tick more activities and name more
colleagues (Cliff's $\delta$ 0.41 and 0.36), and with 134 answers the three
cannot be told apart.

\section{Survey details}
\label{app:survey}

\paragraph{The survey model} The model is a least-squares regression of
$\operatorname{asinh}(\text{hours saved})$ on seven predictors, fitted to the
134 technical coding-agent evaluations (127 respondents) with a
numeric hours answer and every predictor observed. The predictors are
whether the agent enabled
work in the past month that would otherwise have been skipped; the number of
non-coding activity types ticked; whether coding was among the activities;
whether the week's value was only experiments so far; $\log(1 + {}$the number
of colleagues otherwise needed$)$; whether the tool is Agent A (vs.\ Agent
B); and whether
the respondent is in the division with the most respondents in the
survey, which also has the most engineers in our telemetry. All predictors
are standardized. Intervals are 95\% percentiles of
3,000 bootstrap draws that resample respondents, keeping both evaluations of a
respondent together. $R^2$ is 0.43.

\paragraph{The hours measure} Self-reported time savings or losses correlate
with the overall productivity rating at $\rho = 0.57$ and with the ratings
of efficiency in completing routine and complex tasks at 0.56, and with
cognitive load and information-finding at about 0.1; the first two provide
some evidence of convergent validity for the self-reported time savings
measure.

\end{document}